\documentclass[11pt]{article}
\newsavebox{\PSLASH}
\sbox{\PSLASH}{$p$\hspace{-1.8mm}/}

\begin{document}
\title{\large \bf On matching LTB and Vaidya spacetimes through a null hypersurface}
\author{ S. Khakshournia
\footnote{Email address: skhakshour@aeoi.org.ir}
\\
\\
Nuclear Science and Technology Research Institute (NSTRI), Tehran, Iran\\
\\
\\
}\maketitle
\[\]

\begin{abstract}

In this work the matching of a LTB interior solution representing
dust matter to the Vaidya exterior solution describing null fluid
through a null hypersurface is studied. Different cases in which
one is able to smoothly match these two solutions to Einstein
equations along a null hypesurface are discussed.
\end{abstract}
\hspace{1.5cm}
\\
\\

\section{Introduction}

Matching of two exact solutions to the Einstein's equations along
a null hypersurface , although well formulated now \cite{Bar1},
has not had many applications yet. One already knows from the
original Penrose work \cite{Penros} how interesting such matchings
can be: a simple matching of two Minkowski spacetimes may lead to
gravitational impulsive or shock waves across the null boundary.
There are, however, not many examples of such matchings within
general relativity (see \cite{hogen} and references therein). \\
Lemos \cite{lemos1} and Hellaby \cite{Hellabi} have shown that the
Vaidya spacetime which describes incoherent null radiation moving
along null geodesics, can be obtained mathematically from the
Lemaitre-Tolman-Bondi (LTB) spacetime, which represents an
inhomogeneous distribution of dust fluid following timelike
geodesics, by taking the limit of the function
$E(r)\longrightarrow \infty$ in the LTB solution (see Eq.
(\ref{metricLTB}) below). In this case, the two metrics represent
the same collapse and their naked singularities are of the same
nature. Recently, Gao and Lemos \cite{lemos2} studied the
possibility of matching these two solutions in one single
spacetime through a null hyperserface. In this way, by
construction of a continuous coordinate system which is comoving
with both the LTB and Vaidya observers they showed that the
requirement for a smooth matching implies the divergence of $E(r)$
at the place of the null hypersurface. The authors also
demonstrated that the spacetime metric can be at least $C^{1}$ and
the energy- momentum tensor is
continuous across the null hypersurface.\\
In this note we use the Barrab\`{e}s-Israel (BI) null shell
formalism \cite{Bar1} to investigate the smooth matching for the
configuration studied in Ref. \cite{lemos2} and to find the
matching conditions.\\
\textit{Conventions.}  Natural geometrized units, in which $G=c=1$
are used throughout the paper. The null hypersurface is denoted by
$\Sigma$. The symbol $|_{\Sigma}$ means "evaluated on the null
hypersurface ". We use square brackets [F] to denote the jump of
any quantity F across $\Sigma$. Latin indices range over the
intrinsic coordinates of $\Sigma$ denoted by $\xi^{a}$, and Greek
indices over the coordinates of the 4-manifolds.

\section{Matching Conditions}

We choose the LTB metric to be written in the synchronous comoving
coordinates in the form \cite{Bondi}
\begin{equation}\label{metricLTB}
ds^{2}_{-}=-dt^{2}+\frac{{R'}
^{2}}{1+E(r)}dr^{2}+R^{2}(t,r)(d\theta^{2} +\sin^{2} \theta
d\varphi^{2}),
\end{equation}
where the overdot and prime denote partial differentiation with
respect to $t$ and $r$, respectively, and $E(r)$ is an arbitrary
real function of $r$ such that $E(r)>-1$. Then the corresponding
Einstein field equations turn out to be
\begin{eqnarray}\label{field}
\dot{R}^{2}(t,r)&=&E(r)+\frac{2M(r)}{R} ,\\
\label{field2}
\hspace*{0.6cm}4\pi\rho_{L}(t,r)&=&\frac{M'(r)}{R^{2}R'},
\end{eqnarray}
where $\rho_{L}$ is the energy density and $M(r)$ is another
arbitrary function interpreted as the effective gravitational mass
within $r$. We then take an incoming Vaidya spacetime described by
the following metric \cite{vaidya1}
\begin{equation}\label{Vaidya}
ds^{2}_{+}=-\left(1-\frac{2M(v)}{R}\right) dv^{2}+2dvdR+
R^{2}(d\theta^{2}+\sin^{2} \theta d\varphi^{2}),
\end{equation}
where $M(v)>0$ is an arbitrary function of the ingoing null
coordinate $v$, representing the mass accreted at time $v$. The
energy momentum tensor for the metric (\ref{Vaidya}) is of pure
radiation type
\begin{equation}\label{tmuenue}
T_{\mu\nu}=\frac{1}{4\pi R^{2}}\frac{dM(v)}{dv}l_{\mu}l_{\nu},
\hspace{2cm}l_{\mu}=-\delta_{\mu}^{v},\hspace{2cm}l_{\mu}l^{\mu}=0,
\end{equation}
where one can show the identification $dM(v)/dv=4\pi R^2
\rho_{R}$, with $\rho_{R}$ denotes the radiation density. To glue
the interior LTB inhomogeneous region to the exterior Vaidya
spacetime along the null hypersurface $\Sigma$ we need to have
\begin{equation}\label{trans1}
r=r(t),\hspace{1cm}\frac{dr}{dt}=
\frac{\sqrt{1+E}}{R'(t,r)}\big|_{\Sigma},
\end{equation}
in the minus coordinates, while in the plus coordinates, the
hypersurfaces $v=constant$ turn out to be null . Now, the
requirement for the continuity of the induced metric on $\Sigma$
yields the following matching condition
\begin{eqnarray}\label{intrinmatch}
R(t,r)\stackrel{\Sigma}{=}R,
\end{eqnarray}
where $\stackrel{\Sigma}{=}$ means that the equality must be
evaluated on $\Sigma$. For further applications, we note that
differentiation of (\ref{intrinmatch}) on $\Sigma$ leads to
\begin{equation}\label{junctioncond}
\dot{R}dt+R'dr\stackrel{\Sigma}{=}dR.
\end{equation}
Taking $\xi ^{a}=(\lambda,\theta,\varphi)$ with $a=1,2,3$ as the
intrinsic coordinates on $\Sigma$ while identifying $-R$ with the
parameter $\lambda$ on the null generators of the hypersurface we
calculate the tangent basis vectors $e_{a}=\partial/\partial \xi
^{a}$ on both sides of $\Sigma$. Using  Eqs. (\ref{trans1}) and
(\ref{junctioncond}), we get
\begin{eqnarray}\label{verbien}
e^{\mu}_{\lambda}|_{+}&=&\left(0,-1,0,0\right)
\big|_{\Sigma},\hspace*{1.4cm}
e^{\mu}_{\theta}|_{+}=\delta^{\mu}_{\theta},
\hspace*{1.6cm}e^{\mu}_{\varphi}|_{+}=\delta^{\mu}_{\varphi},\\
e^{\mu}_{\lambda}|_{-}&=&\frac{-1}{\dot{R}+\sqrt{1+E}}
\left(1,\frac{\sqrt{1+E}}{R'},0,0\right)\big|_{\Sigma},
\hspace*{0.6cm}e^{\mu}_{\theta}|_{-}=\delta^{\mu}_{\theta},
\hspace*{0.6cm}e^{\mu}_{\varphi}|_{-}=\delta^{\mu}_{\varphi}.
\end{eqnarray}
Choosing the tangent-normal vector $n^{\mu}$ to coincide with the
tangent basis vector associated with the parameter $\lambda$, so
that $n^{\mu} = e^{\mu}_{\lambda}$, we make sure of generating the
null hypersurface $\Sigma$ by the geodesic integral curves of the
future pointing null vector field
$\frac{\partial}{\partial\lambda}$. We may then complete the basis
by a transverse null vector $N^{\mu}$ uniquely defined by the four
conditions $n_{\mu}N^{\mu} = -1$, $N_{\mu}e^{\mu}_{A} = 0$ $(A =
\theta,\varphi)$, and $N_{\mu}N^{\mu} = 0$ \cite{Bar1}. We find
\begin{eqnarray}\label{normaltrans}
N_{\mu}|_{-}&=&\frac{1}{2}(\dot{R}+\sqrt{1+E})\left(1,\frac{R'}{\sqrt{1+E}},0,0\right)\big|_{\Sigma},\\
N_{\mu}|_{+}&=&\left(\frac{-1}{2}\left(1-\frac{2M(v)}{R}\right),1,0,0\right)\big|_{\Sigma}.
\end{eqnarray}

The final matching conditions are formulated in terms of the jump
in the transverse extrinsic curvature. Using the definition ${\cal
K}_{ab} = e^{\mu}_{a}e^{\nu}_{b}\nabla_{\mu}N_{\nu}$ \cite{Bar1},
we now compute the components of the transverse extrinsic
curvature tensor on both sides of $\Sigma$. Its non-vanishing
components on the minus side are found as
\begin{equation}\label{Ktetteta1}
{\cal K}_{\theta\theta}|_{-}=\sin^{-2}\theta{\cal
K}_{\varphi\varphi}|_{-}=\frac{R}{2}\left(1-\frac{2M(r)}{R}\right)\big|_{\Sigma},
\end{equation}
\begin{equation}\label{Krr1}
{\cal K}_{\lambda\lambda}|_{-}=\frac{-M'(r)
}{RR'(\dot{R}+\sqrt{1+E})^{2}}\big|_{\Sigma}.
\end{equation}
The corresponding non-vanishing components on the plus side are
\begin{equation}\label{Ktetteta2}
{\cal K}_{\theta\theta}|_{+}=\sin^{-2}\theta{\cal
K}_{\varphi\varphi}|_{+}=\frac{R}{2}\left(1-\frac{2M(v)}{R}\right)\big|_{\Sigma}.
\end{equation}
Now, the requirement for the continuity of the different
components of the transverse extrinsic curvature tensor across the
null hypersurface $\Sigma$ gives
\begin{equation}\label{junc1}
M(r)\stackrel{\Sigma}{=} M(v),
\end{equation}
\begin{equation}\label{junc2}
\frac{M'(r) }{R'(\dot{R}+\sqrt{1+E})^{2}}\big|_{\Sigma}=0,
\end{equation}
where we have used  Eqs. (\ref{Ktetteta1}-\ref{Ktetteta2}). The
condition (\ref{junc1}) simply expresses that the total
gravitational mass as seen from the exterior must coincide with
that seen from the interior on the hypersurface, as would be
expected . Furthermore, from the condition (\ref{junc2}) we see
that continuity of the $\lambda\lambda$ component of the
transverse curvature tensor over $\Sigma$ requires that one of the
following cases be satisfied:\\
$(i)$ The function $E(r)$ attains the unlimited value (i.e.,
$E(r)\longrightarrow \infty$) on the hypersurface $\Sigma$. Given
the following limiting forms on the hypersurface in a LTB
collapsing model as studied in \cite{lemos2}

\begin{eqnarray}\label{limitform}
R&\longrightarrow& \sqrt{E}(a-t),\nonumber\\
\dot{R}&\longrightarrow& -\sqrt{E},\\
R'&\longrightarrow &\frac{RE'}{2E}+a'\sqrt{E},\nonumber
\end{eqnarray}
where $a(r)$ is an arbitrary function, we can see that the
denominator in the matching condition (\ref{junc2}) goes to
infinity so as to for the nonzero values of the numerator the
condition (\ref{junc2}) will be satisfied.\\
$(ii)$ $M'(r)$ approaches zero on the hypersurface $\Sigma$
while the function $E(r)$ remains finite.\\
In both cases $(i)$ and $(ii)$ from Eq. (\ref{field2}) we see that
the energy density of LTB interior region $\rho_{L}$ must go to
zero on the hypersurface $\Sigma$, otherwise there will be a
nonzero isotropic surface pressure given by \cite{Bar1}
\begin{equation}\label{pressure}
p = -[{\cal K}_{\lambda\lambda}]= \frac{-4\pi
R\rho_{L}}{(\dot{R}+\sqrt{1+E})^{2}}\big|_{\Sigma},
\end{equation}
signaling the presence of a thin distribution of matter on the
hypersurface $\Sigma$ with the surface quantity (\ref{pressure})
which would be better described as a tension.\\
In the case of smooth matching of the two solution the jump in the
component $T_{\mu\nu}n^{\mu}n^{\nu}$ of the energy momentum tensor
is zero and in addition, it follows that $\lambda$ is an affine
parameter on the both sides of $\Sigma$. \\
The Kretschmann scalar
$K=R^{\mu\nu\sigma\rho}R_{\mu\nu\sigma\rho}$ is computed for the
LTB spacetime as \cite{lemos2}
\begin{equation}\label{K1}
K|_{-}=\frac{48M^{2}}{R^{6}}-\frac{32MM'}{R^{5}R'}+\frac{12{M'}
^{2}}{R^{4}R'^{2}},
\end{equation}
and for the Vaidya spacetime
\begin{equation}\label{K2}
K|_{+}=\frac{48M^2(v)}{R^{6}}.
\end{equation}
By virtue of the matching conditions (\ref{junc1}) and
(\ref{junc2}) together with the limiting form $R'\longrightarrow
\infty$ related to the case $(i)$ one can easily show the
continuity of the Kretschmann scalar over the null hypesurface for
the both cases $(i)$ and $(ii)$.

\section{Conclusion}

Applying the Barrab\`{e}s-Israel null shell formalism we have
examined the matching of a LTB interior region to the Vaidya
exterior spacetime along a null hypersurface. We have shown that
in the context of BI formalism the limit of interest in which E(r)
diverges on the null hypersurface as considered in \cite{lemos2}
stems from the requirement for the continuity of the components of
the second fundamental form across the hypersurface as needed for
the smooth matching of the two spacetimes. Furthermore, we
conclude that one can glue these two solutions to the Einstein
equations smoothly through a null hypersurface even for the finite
values of the function E(r) provided that the function
$M'(r)\longrightarrow 0$ on $\Sigma$. In any case, the energy
density of LTB interior region $\rho_{L}$ goes to zero on the
hypersurface $\Sigma$.
\subsection*{Acknowledgment}
This work was motivated by a stimulating correspondence with
Professor J. P. S. Lemos.

\end{document}